\def\year{2019}\relax
\documentclass[letterpaper]{article} 
\usepackage{aaai19}  
\usepackage{times}  
\usepackage{helvet} 
\usepackage{courier}  
\usepackage[hyphens]{url}  
\usepackage{graphicx} 
\def\UrlFont{\rm}  
\usepackage{graphicx}  
\usepackage{amsmath}
\newcommand{\newcite}[1]{\citeauthor{#1} \shortcite{#1}}

\newcommand{\Sref}[1]{\S\ref{#1}}

\newcommand{\Fref}[1]{Figure~\ref{#1}}

\newcommand{\Tref}[1]{Table~\ref{#1}}

\title{Contextual Affective Analysis: \\ A Case Study of People Portrayals in Online \#MeToo Stories }
\author{Anjalie Field, Gayatri Bhat, Yulia Tsvetkov\\ 
Language Technologies Institute\\
Carnegie Mellon University\\
\{anjalief, ytsvetko\}@cs.cmu.edu 
}

\begin{document}

\maketitle

\begin{abstract}
In October 2017, numerous women accused producer Harvey Weinstein of sexual harassment. Their stories encouraged other women to voice allegations of sexual harassment against many high profile men, including politicians, actors, and producers. These events are broadly referred to as the \#MeToo movement, named for the use of the hashtag ``\#metoo'' on social media platforms like Twitter and Facebook. The movement has widely been referred to as ``empowering'' because it has amplified the voices of previously unheard women over those of traditionally powerful men. In this work, we investigate dynamics of sentiment, power and agency in online media coverage of these events. Using a corpus of online media articles about the \#MeToo movement, we present a \textit{contextual affective analysis}---an entity-centric approach that uses contextualized lexicons to examine how people are portrayed in media articles. We show that while these articles are sympathetic towards women who have experienced sexual harassment, they consistently present men as most powerful, even after sexual assault allegations. While we focus on media coverage of the \#MeToo movement, our method for contextual affective analysis readily generalizes to other domains.\footnote{We provide code and public data at \url{https://github.com/anjalief/metoo_icwsm2019}} 
\end{abstract}

\section{Introduction}

In 2006, Tarana Burke founded the \#MeToo movement, aiming to promote hope and solidarity among women who have experienced sexual assault \cite{metoo_origins}.
In October 2017, following waves of sexual harassment accusations against producer Harvey Weinstein, actress Alyssa Milano posted a tweet with the hashtag \#MeToo and encouraged others to do the same. Her message initiated a widespread movement, calling attention to the prevalence of sexual harassment and encouraging women to share their stories.

Tarana Burke has described her primary goal in founding the movement as ``empowerment through empathy.''\footnote{\url{https://metoomvmt.org/}} However, mainstream media outlets vary in their coverage of these recent events, to the extent that some outlets accuse others of misappropriating the movement. For instance, in January 2018, \url{Babe.net} published an article written by Katie Way, describing the interaction between anonymous `Grace' and famous comedian Aziz Ansari \cite{babe}. The article sparked not only instant support for Grace, but also instant backlash criticizing Grace's lack of agency: ``The single most distressing thing to me about this story is that the only person with any agency in the story seems to be Aziz Ansari'' \cite{nyt_grace}. One widely circulated article, written by Caitlin Flanagan and published in The Atlantic, strongly criticized Way's article and questioned whether modern conventions prepare women to fight back against potential abusers \cite{atlantic}.

The manner in which accounts of sexual harassment portray the people involved affects both the audience's reaction to the story and the way people involved in these incidents interpret or cope with their experiences \cite{spry1995absence}. In this work, we use natural language processing (NLP) techniques to analyze online media coverage of the \#MeToo movement. In a \emph{people-centric} approach, we analyze narratives that include individuals directly or indirectly involved in the movement: victims, perpetrators, influential commenters, reporters, etc. Unlike prior work focused on social media \cite{ribeiro2018media,rho2018fostering}, our work examines the prominent role that more traditional outlets and journalists continue to have in the modern-era online media landscape.

In order to structure our approach, we draw from social psychology research, which has identified 3 primary affect dimensions:  \emph{Potency} (strength vs. weakness), \emph{Valence} (goodness vs. badness), and \emph{Activity} (liveliness versus torpidity) \cite{osgood1957measurement,russell1980circumplex,russell2003core}. Exact terminology for these terms has varied across studies. For consistency with prior work in NLP, we refer to them as \textbf{power}, \textbf{sentiment}, and \textbf{agency}, respectively \cite{SapConnotationFilms,RashkinConnotationInvestigation}. In the context of the \#MeToo movement, these dimensions tie closely to the concept of ``empowerment through empathy.''

The crux of our method is in developing contextualized, entity-centric connotation frames, where polarity scores are generated for words in context. We generate token-level sentiment, power, and agency lexicons by combining contextual ELMo embeddings \cite{peters2018deep} with (uncontextualized) connotation frames \cite{RashkinConnotationInvestigation,SapConnotationFilms} and use supervised learning to propagate annotations to unlabeled data in our \#MeToo corpus. Following prior work, we first evaluate these models over held-out subsets of the connotation frame annotations. We then evaluate the specifics of our method, namely contextualization and entity scoring, through manual annotations.

\begin{figure*}[ht]
  \includegraphics[width=\linewidth]{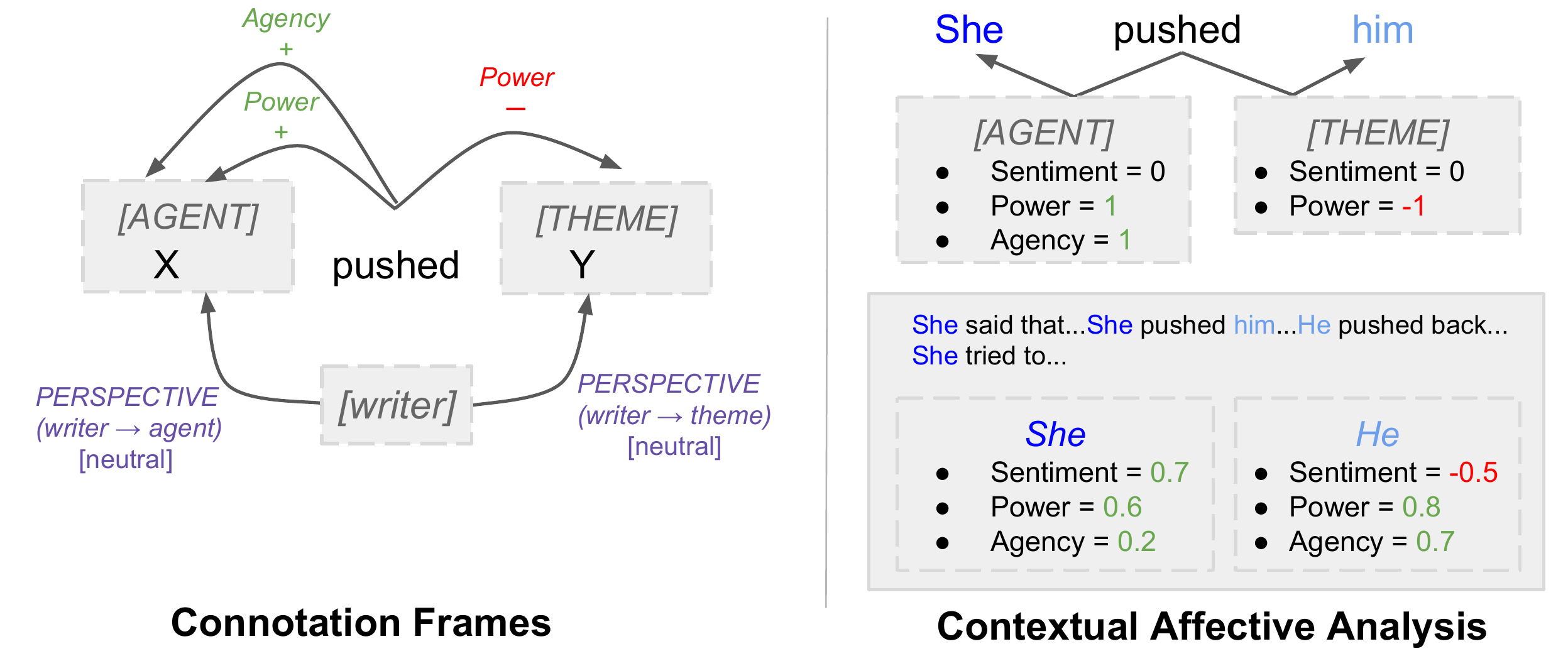}
  \caption{Left, we show off-the-shelf connotation frame annotations  \cite{RashkinConnotationInvestigation,SapConnotationFilms} for the verb ``push''. Right, we show the proposed adaptation. We adapt connotation frames from a verb-centric formalism to an entity-centric formalism, transferring scores from verbs to entities using a context-aware approach (top right). We then aggregate contextualized scores over all mentions of entities in a corpus (bottom right). This new approach---\textit{contextual affective analysis}---enables us to obtain sentiment, power, and agency scores for entities in unannotated corpora and conduct extensive analyses of people portrayals in narratives, which we exemplify on \#MeToo data.}
  \label{fig:push}
\end{figure*}

We ultimately use these contextualized connotation frames to generate sentiment, power, and agency scores for entities in news articles related to the \#MeToo movement. We find that while the media generally portrays women revealing stories of harassment positively, these women are often not portrayed as having high power or agency, which threatens to undermine the goals of the movement. To the best of our knowledge, this is the first work to introduce \textit{contextual affective analysis}, a method that enables nuanced, fine-grained, and directed analyses of affective social meanings in narratives. 

\section{Background}
We motivate the development of contextual affective analysis as a people-centric approach to analyzing narratives. Entity-centric models, which focus on people or characters rather than plot or events, have become increasingly common in NLP \cite{bamman2015people}. However, most approaches rely on unsupervised models \cite{iyyer2016feuding,chambers2009unsupervised,bamman2013learning,card2016analyzing}, which can capture high-level patterns but are difficult to interpret and do not target specific dimensions.

In contrast, we propose an interpretable approach that focuses on power, sentiment, and agency. These dimensions are considered both distinct and exhaustive in capturing affective meaning, in that all 3 dimensions are needed, and no additional dimensions are needed; other affective concepts, such as anger or joy, are thought to decompose into these three dimensions \cite{russell1980circumplex,russell2003core}. Furthermore, these dimensions form the basis of \textit{affective control theory}, a social psychological model which broadly addresses how people respond emotionally to events and how they attribute qualities to themselves and others \cite{heise1979understanding,heise2007expressive,robinson2006affect}. Affective control theory has served as a model for stereotype detection in NLP \cite{joseph2017girls}.

Furthermore, while automated sentiment analysis has spanned many areas \cite{pang2008opinion,liu2012sentiment}\footnote{\url{http://alt.qcri.org/semeval2016/index.php?id=tasks}} analysis of power has been almost entirely limited to a dialog setting: how does person A talk to a higher-powered person B? \cite{gilbert2012phrases,Prabhakaran2017DialogPower,danescu2012echoes}. Here, we focus on a \emph{narrative} setting: does the journalist portray person A or person B as more powerful?

In order to develop an interpretable analysis that focuses on sentiment, power, and agency in narrative, we draw from existing literature on \textit{connotation frames}: sets of verbs annotated according to what they imply about semantically dependent entities. Connotation frames, first introduced by \newcite{RashkinConnotationInvestigation}, provide a framework for analyzing nuanced dimensions in text by combining polarity annotations with frame semantics \cite{fillmore1982frame}. We visualize connotation frames in \Fref{fig:push} on the left. More specifically, verbs are annotated across various dimensions and perspectives, so that a verb might elicit a positive sentiment for its subject (i.e. sympathy) but imply a negative effect for its object. We target power, agency, and sentiment of entities through pre-collected sets of verbs that have been annotated for these traits:
\begin{itemize}
        \item Perspective($writer \rightarrow agent$) -- Does the writer portray the agent (or subject) of the verb as positive or negative?
        \item Perspective($writer \rightarrow theme$) -- Does the writer portray the theme (or object) of the verb as positive or negative?
        \item Power -- does the verb imply that the theme or the agent has power?
       \item Agency -- does the verb imply that the subject has positive agency or negative agency?
\end{itemize}

 For clarity, we refer to Perspective($writer \rightarrow agent$) as Sentiment($agent$) and Perspective($writer \rightarrow theme$) as Sentiment($theme$) throughout this paper.
 
 These dimensions often differ for the same verb. For example, in the sentence: ``She amuses him,''  the verb ``amuses'' connotes that \textit{she} has high agency, but \textit{he} has higher power than \textit{she}. \newcite{RashkinConnotationInvestigation} present a set of verbs annotated for sentiment, while \newcite{SapConnotationFilms} present a set of verbs annotated for power and agency. However, these lexicons are not extensive enough to facilitate corpus analysis without further refinements. First, they contain only a limited set of verbs, so a given corpus may contain many verbs that are not annotated. Furthermore, verbs are annotated in synthetic context (e.g., ``X amuses Y''), 
rather than using real world examples. Finally, each verb is annotated with a single score for each dimension, but in practice, verbs can have different connotations in different contexts.

Consider two instances of the verb ``deserve'': 

\begin{enumerate}
    \item The hero deserves appellation
    \item The boy deserves punishment
\end{enumerate}

In the first instance, annotators rate the writer's perspective towards the agent (``hero'') as positive, while in the second instance, annotators rate the writer's perspective towards the agent (``boy'') as negative \cite{RashkinConnotationInvestigation}. We can find numerous similar examples in articles related to the \#MeToo movement, i.e. \textit{She pushed him away} vs \textit{Will one part of the movement's legacy be to push society to find the right words to describe it all?}.

The uncontextualized annotations presented by \newcite{RashkinConnotationInvestigation} and \newcite{SapConnotationFilms} serve as starting points for more in-depth analysis. We build uncontextualized features for verbs to match the uncontextualized annotations, and then use supervised learning to extend the uncontextualized annotations to verbs in context, thus learning contextualized annotations.

Our work extends the concept of domain-specific lexicons: that words have different connotations in different situations. For instance, over the last century, the word ``lean'' has lost its negative association with ``weakness'' and instead become positively associated with concepts like ``fitness.'' These changes in meaning have motivated research on inducing domain-specific lexicons \cite{hamilton2016inducing}. Using contextual embeddings \cite{peters2018deep}, we extend this concept by introducing \emph{context-specific} lexicons: we induce annotations for words in context, rather than just words in domain. To the best of our knowledge, this is the first work to propose contextualized affective lexicons. Our overall methodology uses these contextualized lexicons to obtain power, sentiment, and agency scores for entities in contextual affective analysis.

\section{Methodology}
\label{sec:methodology}
In the proposed contextual affective analysis, our primary goal is to analyze how people are portrayed in narratives. To obtain sentiment, power, and agency scores for people in the context of a narrative (a sentence, paragraph, article, or an outlet), we adapt the connotation frames reviewed above as follows. First, since connotation frames are verb-centric---rather than people-centric---we define a mapping from a verb in a sentence to people that are syntactic arguments of the verb. We visualize this mapping in Figure \ref{fig:push}. Second, since only a small subset of verbs ($<$30\%) in our corpus is included in the annotations crowdsourced by \newcite{RashkinConnotationInvestigation} and \newcite{SapConnotationFilms}, we devise a lexicon induction method to annotate unlabeled verbs using the existing seed of annotations. To contextualize the analysis, the lexicon induction procedure uses contextual ELMo embeddings \cite{peters2018deep}. Contextual embeddings are a form of distributed word representations that incorporate surrounding context words. Thus, using the above example, ELMo embeddings provide different representations for ``push'' in ``She pushed him away'' and for ``push'' in ``Will one part of the movement's legacy be to push...'' In what follows, we detail the components of our methodology. All of our code is publicly available.

We use connotation frames, which are annotations on verbs, to obtain affective scores on people (the agent or theme of these verbs). In our running example ``She pushes him away,'' ``She'' is the agent  and ``him'' is the theme. Given a verb $V$, the verb's agent $A$, the verb's theme $T$, and a set of connotation frame annotations over $V$ (e.g., $V_{Sentiment(agent)} = +1$), we obtain sentiment, power, and agency scores as follows:

\begin{align*}
Sentiment(A) &= V_{Sentiment(agent)}\\
Power(A) &= V_{Power}\\
Agency(A) &= V_{Agency}\\
Sentiment(T) &= V_{Sentiment(theme)}\\
Power(T) &= -V_{Power}\\
\end{align*}

We obtain a sentiment score within a corpus for an entity $E$ by averaging over all $V_{Sentiment(agent)}$ scores where $E$ is the agent of $V$ and all $V_{Sentiment(theme)}$ scores where $E$ is the theme of $V$. We compute agency and power scores in the same way. Thus, the final entity scores are not merely direct mappings from verb annotations, but an aggregation of all such mappings over the corpus of entity mentions.

We obtain these verb scores by taking a supervised approach to labeling verbs in context with sentiment ($V_{Sentiment(agent/theme)}$~$\in\{-1, 0, +1\}$), power, ($V_{Power}$~$\in\{-1, 0, +1\}$), and agency ($V_{Agency}$~$\in\{-1, 0, +1\}$).

For a given verb in our training set $V$, we assume that $V$ occurs $n$ times in our corpus, and enumerate these occurrences as $V_1...V_n$. For each $V_i$, we compute the contextualized ELMo embedding $\mathbf{e}_i$. We then ``decontextualize'' these embeddings by averaging over $\mathbf{e}_1 \ldots \mathbf{e}_n$ to obtain a single feature representation $\mathbf{e}$. We consider these decontextualized embeddings to be representative of the off-the-shelf uncontextualized lexicons, and we use them as training features in a supervised classifier.

Then, for a given verb in our corpus $T$, for each instance where $T$ occurs in our corpus, we use its contextualized ELMo embedding $\mathbf{e}_i$ as a feature to predict an annotation score for $T_i$. In particular, we use logistic regression with re-weighting of samples to maximize for the best average F1 score over a dev set.\footnote{We experimented with other supervised and semi-supervised methods common in lexicon induction including graph-based semi-supervised label propagation and random walk-based propagation \cite{goldberg2006seeing,hamilton2016inducing}, but found that logistic regression outperformed these methods on all frames.}

For evaluation, in order to compare our results against existing annotations, we use two methods to obtain uncontextualized annotations from the learned $T_i$ scores for verbs in our test sets. In the first, which we refer to as \textit{type-level}, we average all of the token-level embeddings ($\mathbf{e}_i$) in the test data in the same way as in the training data, and we learn a single annotation for each verb $T$, rather than learning contextualized annotations. This approach is most similar to prior work. In the second, which we refer to as \textit{token-level}, we predict a separate score for each token-level embedding as described above, and we then take a majority vote over scores to obtain an overall score for each verb.

\section{Experimental Setup}

\begin{table}
\begin{tabular}{lcc}
                  &  \textbf{Lexicon Size} & \textbf{Training Set Size} \\
                  
\textbf{Sentiment} & 948       & 300    \\
\textbf{Power}  & 1,714       & 571     \\ 
\textbf{Agency}  & 2,104       & 701    \\

\end{tabular}
\caption{Annotated Lexicon Statistics}
\label{tab:lex_stats}
\end{table}

\paragraph{Lexicons}

 We provide basic statistics of the annotated lexicons in Table~\ref{tab:lex_stats}. The sentiment frame annotations are reported averages over annotations from 15 crowd-workers. We ternerize these annotations using the same cut-offs as \newcite{RashkinConnotationInvestigation}: Negative: $[-1,-0.25)$, Neutral: $[-0.25, 0.25]$ and Positive: $(0.25, 1]$. The power frame annotations are formatted as [power\_agent, power\_equal, power\_theme], which we map to $[1, 0, -1]$. Similarly, the agency verbs are formatted as [agency\_positive, agency\_equal, agency\_negative], which we map to $[1, 0, -1]$.

\paragraph{Corpora}
We gathered a corpus of articles related to the \#MeToo movement by first collecting a list of URLs of articles that contain the word ``metoo'' using an API that searches for articles from over 30,000 news sources.\footnote{\url{https://newsapi.org/}} Over two separate queries, we gathered URLs from November 2, 2017 to January 31, 2018 and from February 28, 2018 to May 29, 2018. Next, we used Newspaper3k to obtain the full text of each article.\footnote{\url{http://newspaper.readthedocs.io/en/latest/}}

We then discarded any pages that returned 404 errors, any URLs containing the word ``video,'' and any non-English articles, identifying languages using the Python package langdetect.\footnote{\url{https://pypi.org/project/langdetect/}} Finally, we removed duplicate articles by converting each article into a bag-of-words vector and computing the cosine distance between every pair of vectors. If the distance between 2 articles was less than a threshold (0.011, identified by manually examining random samples), we discarded the more recent article. Our final data set consists of 27,602 articles across 1,576 outlets, published between November 2, 2017 and May 29, 2018.

Our data collection method includes any articles that mention the word \#MeToo, which includes articles focused on other events that only mention the movement in passing. However, we believe these articles are still relevant for our analysis, as mentioning any entities alongside the movement implicitly associates them with these events, and as discussed in \Sref{sec:analysis}, people not directly involved in events can become important entities in the movement.

\paragraph{Preprocessing} We tokenize and sentence-split our corpus using the Stanford NLP pipeline. In generating 1024-dimensional ELMo embeddings, we only take embeddings for verbs, performing stemming and POS tagging using spaCy\footnote{https://spacy.io/}, and we keep only the 2nd (middle) ELMo embedding layer. In generating entity scores, we use dependency parsing to identify agents and themes: we consider an entity $E$ to be an agent of $V$ if it is the verb's subject. We consider an entity $E$ to be a theme of $V$ if it is the verb's object or passive subject (``nsubjpass''). We use the Stanford NLP pipeline for dependency parsing, named entity recognition, and co-reference resolution. We find a total of 3,132,389 entity-verb pairs across the corpus, which form the basis of our analysis.

\section{Evaluation}

\begin{table}
\begin{tabular}{lll|ll}
\textbf{}      & \textbf{Aspect} & \textbf{Frame} & \textbf{Type} & \textbf{Token} \\
               &                 &                &                     &           \\
\multicolumn{5}{l}{\textbf{Sentiment ($theme$)}}                \\
Accuracy       & 67.56           & 67.56          & 63.33               & 66.67     \\
Macro F1       & 56.18           & 56.18          & 55.63               & 51.90          \\
               &                 &                &                     &                 \\
\multicolumn{5}{l}{\textbf{Sentiment ($agent$)}} \\
Accuracy       & 60.54           & 61.87          &  60.00           & 61.33              \\
Macro F1       & 60.72           & 63.07          &  58.26           & 60.23             \\
               &                 &                &                     &                      \\
               
\textbf{}      & \textbf{} & \textbf{Majority} & \textbf{Type} & \textbf{Token} \\
\textbf{}      & \textbf{} & \textbf{Class} & \textbf{} & \textbf{} \\
\multicolumn{5}{l}{\textbf{Power}} \\
Accuracy       & -           & 69.66          & 69.66            & 69.49           \\
Macro F1       & -           & 27.37          & 55.97            & 54.10             \\
               &                 &                &                     &             \\
 \multicolumn{5}{l}{\textbf{Agency}} \\
Accuracy       & -           & 79.14         &  70.30            & 72.74              \\
Macro F1       & -           & 29.45          & 48.79            & 50.14             \\

\end{tabular}
\caption{Accuracy and F1 Score of lexicon expansion for our methods (Type and Token) compared with prior work (Aspect and Frame). Non-trivial F1 scores demonstrate that our feature representations capture meaningful information about sentiment, power, and agency.}
\label{tab:res}
\end{table}

We first evaluate our methods on their ability to predict the labels of off-the-shelf connotation frame lexicons. While this task is not our primary objective, it serves as a sanity-check on our feature representations and allows us to compare our method with prior work. Then, we evaluate the methods in a contextualized setting, assessing their ability to model contextualized verb annotations (\Sref{sec:contextual_evaluation}) and contextualized entity annotations (\Sref{sec:entity_evaluation}) by comparing with human annotators.

We divide the annotations into train, dev, and test; for sentiment, we use the same data splits as \newcite{RashkinConnotationInvestigation}; for power and agency, we randomly divide the lexicons into subsets of equal size. In order to compare the contextualized scores generated by our method with the off-the-shelf annotations, we aggregate the contextualized annotations into uncontextualized annotations for the verbs in the test set, as described in \Sref{sec:methodology}.

Table \ref{tab:res} reports results. For comparison, we show the Aspect-Level and Frame-Level models presented by \newcite{RashkinConnotationInvestigation} over the sentiment annotations. Our type-level logistic regression is essentially identical to the aspect-level model, the primary difference being our use of ELMo embeddings. Our results are slightly lower, but generally comparable to the results reported by \newcite{RashkinConnotationInvestigation}; crucially, they are obtained with a model that ultimately allows us to incorporate context. The type-level and token-level aggregation methods perform about the same.

 In the absence of prior work on this task for the power and agency lexicons, we show a majority class baseline. Our methods show a strong improvement over F1 scores for the majority class baseline. As for sentiment, the type-level and token-level methods perform similarly. \Tref{tab:res} generally shows that ELMo embeddings capture meaningful information about power, agency, and sentiment. However, our primary task is not to re-create the word-level annotations in the connotation frame lexicons, but rather to contextualize these lexicons by obtaining instance-level scores over verbs in context. We evaluate these contextualized scores in the following section.

\subsection{Evaluation of Contextualization}
\label{sec:contextual_evaluation}

In this section, we draw from the original annotations used to create the connotation frame lexicons in order to assess the impact of contextualization. The publicized connotation frames consist of a single score for each verb. However, for the sentiment dimensions, these scores were obtained by collecting annotations over verbs in a variety of simple synthetically-generated contexts and averaging annotations across contexts, i.e. collecting 5 annotations each for ``the hero deserves appellation,'' ``the student deserves an opportunity,'' and ``the boy deserves punishment'' and averaging across the 15 annotations \cite{RashkinConnotationInvestigation}. Then, for the sentiment lexicons, we can evaluate our method's ability to provide annotations in context by reverting to the original pre-averaged annotations. (We cannot perform the same evaluation for the power and agency lexicons, because they were created by collecting annotations over verbs without any context, i.e. ``X deserves Y'' \cite{SapConnotationFilms}.)

When we ignore context, meaning we treat all 15 annotations over each verb as annotations over the same sample, the inter-annotator agreement (Krippendorff's alpha) for Sentiment($theme$) is 0.20 and for Sentiment($agent$) is 0.28. However, when we treat each sentence as a separate sample (i.e. measuring agreement in annotations over ``the hero deserves appellation'' separately from annotations over ``the boy deserves punishment''), the agreement rises to 0.34 and 0.40 respectively, a $>40\%$ increase for each trait. The improvement in agreement demonstrates that when annotators disagree about the connotation implied by a verb, it is often because the verb has different connotations in different contexts.

\begin{table}
\begin{tabular}{lccc}
    & \textbf{Verb-level} & \textbf{Sent.-level} & \textbf{Sent.-level} \\
    &  &  & \textbf{training} \\
                  
\textbf{Sentiment} ($t$) & 41.05       & 44.35  & 50.16  \\
\textbf{Sentiment} ($a$) & 51.37       & 52.80  &  54.11  \\ 
\end{tabular}
\caption{F1 scores for using our method to score contextualized annotations. Predicting sentence-level scores outperforms predicting verb-level scores. Best performance is achieved by also using sentence-level training data, but this is unsustainable in practice.}
\label{tab:raw_frames}
\end{table}

We can then evaluate our method for contextualization by using these sentence-level annotations. More specifically, we use the same train, dev, and test splits as before. However, for verbs in the test set, instead of averaging all 15 annotations for each verb into a single score, we only average over annotations on the same sentence. Thus, our gold test data contains separate scores for ``the hero deserves appellation'', ``the student deserves an opportunity'', and ``the boy deserves punishment'', resulting in approximately 3 times as many test points as the test data in \Tref{tab:res}.

\Tref{tab:raw_frames} shows the results of evaluating our method on this contextualized test set. In the first column, Verb-level, our model disregards contextualization and predicts a single score for each verb using type-level aggregation, which is equivalent to the method used in \Tref{tab:res}. The primary difference is that in \Tref{tab:res}, we evaluate over uncontextualized annotations (i.e. a single score for ``deserve''), while in \Tref{tab:raw_frames}, we evaluate over contextualized annotations.

In the second column, Sent.-level, we predict a separate score for each verb in context, rather than using token or type level aggregation over the test data. This column represents our primary method and is the method we use for analysis in \Sref{sec:analysis}. For both traits, this method outperforms the aggregation approach shown in the first column.

In the third column, Sent.-level training, we similarly predict a separate score for each context, but we further treat each context as a separate training sample. Thus we both train and evaluate on contextualized annotations. In the Sent.-level and Verb-level columns, we train on uncontextualized annotations, as described in \Sref{sec:methodology}.

While training on contextualized annotations achieves the best performance, it is difficult to generalize to other data sets. The sentences used for gathering these annotations were created using Google Syntactic N-grams and designed to be short generic sentences \cite{RashkinConnotationInvestigation}. Thus, they are much simpler than real sentences, and we would not expect them to serve as realistic training data in other domains. In order to use these connotation frame lexicons in a new domain, it would be necessary to annotate a new set of sentences, which defeats the usefulness of off-the-shelf lexicons. Instead, we focus on the second column, Sent.-level, as our primary method, since it is an improvement over existing ways of using off-the-shelf lexicons without requiring new annotations for every task. Overall, \Tref{tab:raw_frames} demonstrates the usefulness of contextualization, as both contextualized approaches outperform the uncontextualized approach.

\subsection{Evaluation of Entity Scores}
\label{sec:entity_evaluation}

In the previous sections, we evaluated our methods for generating verb annotations. In this section, we evaluate our methods for transferring verb annotations to entity scores, specifically focusing on power. In order to assess entity scoring, we devised an annotation task in which we asked annotators to read articles and rank entities mentioned. We then compare our entity scores against these annotations.

More specifically, we sampled 30 articles from our corpus that all contain mentions of Aziz Ansari. We then provided 2 annotators with a list of 23 entities extracted from these articles and asked the annotators to read each article in order. After every 5 articles, annotators ranked the listed entities by assigning a 1 to the lowest-powered entity, a 7 to the highest-powered entity, and scaling all other entities in between. In this way, since annotators rerank entities every 5 articles, we maximize the number of annotations we obtain, while minimizing the number of articles that annotators need to read. Furthermore, we ensure that we obtain different power scores for different entities by forcing annotators to use the full range of the ranking scale.

However, we note that this is a very subjective annotation task. Annotators specifically described it as difficult, both in deciding which entities were most powerful and in ranking entities based on the provided articles rather than on outside knowledge. We observed some of this subjectivity in the collected annotations: one annotator consistently ranked abstract entities like ``The New York Times'' as high-powered, while the other annotator consistently ranked these entities as low-powered. Thus, while we present results as an approximation of how well our methods work, we caution that further evaluation is needed.

From the annotations, we have a ranking for each entity for each 5-article step. Hypothesizing that some entities are more subjective to rank than others, we eliminate all samples where the difference in ranking between the two annotators is greater than 2. We are then left with 81 annotations. The correlation between annotators on this set is statistically significant (Spearman's R=0.55, p-value=1.03e-07). In the following analysis, we average the rank assigned by annotators to obtain a score for each entity.

We ultimately evaluate our methods through pairwise comparisons at each 5-article step. For every pair (A, B) of entities at each time step, we evaluate if entity A is scored as more powerful or less powerful than entity B. We discard samples where the entities were ranked as equal. We compare our method against 2 baseline metrics: (1) the frequency of the entity and (2) power scores assigned by the off-the-shelf connotation frames, rather than our contextualized frames.

\begin{table}
\centering
\begin{tabular}{ccc}
\textbf{Off-the-shelf} & \textbf{Frequency} & \textbf{Ours} \\
                  57.1 & 59.1  & 71.4  \\

\end{tabular}
\caption{Accuracy for scoring how powerful entities are, as compared with manual annotations. We calculate accuracy by assessing if the metric correctly answers ``is entity A more powerful than entity B?''. Our method outperforms both baseline metrics.}
\label{tab:power_annotations}
\end{table}

Off-the-shelf connotation frames are limited to a subset of verbs in our corpus. Furthermore, our analysis pipeline is dependent on the named entity extraction and co-reference resolution tools used during pre-processing, which we find miss many entity mentions (for instance, when we manually extract entities in the first 10 articles, we identify 28 entities that occur at least 3 times, while the automated pipeline identifies only 4). For fair comparison between our method and the off-the-shelf lexicons, we discard entities that do not occur with at least 3 off-the-shelf power-annotated verbs, as identified by our preprocessing pipeline. After this filtering, we are left with 49 pairwise comparisons.

\Tref{tab:power_annotations} reports results. Our method outperforms both baselines, correctly identifying the higher-ranked entity 71.4\% of the time. We note that each annotator individually achieves at most 83.7\% accuracy on this task, which suggests an upper limit on the achievable accuracy.

Furthermore, if we perform the same test, eliminating only entities that occur fewer than 3 times in the text, rather than mandating that entities occur with at least 3 off-the-shelf annotated verbs, our method achieves an accuracy of 63.01\% over 73 pairs, while the frequency baseline achieves an accuracy of 53.42\%.

In conducting this analysis, we observed that one of the limitations of the power annotations provided by \newcite{SapConnotationFilms} is that the authors only annotate transitive verbs for power. They hypothesize that a power differential only occurs when an entity (e.g. the agent) has power over another entity (e.g. the theme). However, we do not limit our scoring to transitive verbs, hypothesizing that intransitive verbs can also be indicative of power, even if there is no direct theme. The improved performance of our scoring metric over the off-the-shelf baseline supports this hypothesis.

\section{Analysis of Entity Portrayals in \#MeToo Movement}
\label{sec:analysis}

In this section, we use the affect scores to analyze how people are portrayed in media coverage of the \#MeToo movement. For reference, we provide brief descriptions of the entities mentioned and their connection to the movement in the Appendix. We propose a top-down framework for structuring a people-centric analysis with three primary levels:
\begin{enumerate}
    \item Corpus-level: we examine broad trends in coverage of all common entities across the entire corpus
    \item Role-level: we examine how people in similar roles across separate incidents are portrayed
    \item Incident-level: we restrict our analysis to people involved in a specific incident
\end{enumerate}

We present here only a subset of possible analyses. While we focus on the \#MeToo movement, our methodology readily generalizes to other research questions and corpora, such as analyzing news coverage of scientific publications \cite{maclaughlin2018predicting}.

\subsection{Corpus-Level}
By examining portrayals at a corpus level, we can assess the overall media coverage of the \#MeToo movement. Whom does the media portray as sympathetic? Did media coverage of events empower individuals?

\begin{table}
\begin{tabular}{ll}
\textbf{Most Positive} & \textbf{Most Negative} \\
Kara Swisher           & Bill Cosby             \\
Tarana Burke           & Harvey Weinstein       \\
Meghan Markle          & Eric Schneiderman    \\
Frances McDormand      & Kevin Spacey \\
Oprah Winfrey          &  Ryan Seacrest         
\end{tabular}
\caption{The most positively portrayed entities consist primarily of 3rd party commentators on events. The most negatively portrayed entities consist of men accused of sexual harassment.}
\label{tab:corpus_sentiment}
\end{table}

We examine these questions by computing sentiment, power, and agency scores for the 100 most frequent proper nouns across the corpus, shown in part in Tables~\ref{tab:corpus_sentiment}--\ref{tab:agency_all}. For brevity, we omit redundant entities (i.e. Donald Trump and President Donald Trump). Table \ref{tab:corpus_sentiment} shows the five most positively and negatively portrayed entities. Unsurprisingly, the most negatively portrayed entities all consist of men accused of sexual harassment, lead by Bill Cosby, the first man actually convicted in court following the wave of accusations in the movement. However, the most positively portrayed entities consist not of women voicing accusations or of men facing them, but rather of 3rd party commentators, who were outspoken in their support of the movement. None of these entities were directly involved in cases that arose out of the \#MeToo movement, but all of them made widely-circulated comments in support of the accusers.

\begin{table}
\begin{tabular}{ll}
\textbf{Highest Power} & \textbf{Lowest Power} \\
The \#MeToo movement     & Kevin Spacey      \\
Judge Steven O'Neill    & Andrea Constand   \\
The New York Times      & Uma Thurman       \\
Congress                & Dylan Farrow      \\
Facebook                & Leeann Tweeden    \\
Twitter                 &                   \\
Eric Schneiderman       &                   \\
Donald Trump            &                   \\
\end{tabular}
\caption{The entities portrayed as most powerful consist of men and abstract institutions. The lowest powered entities consist of primarily women.}
\label{tab:power_all}
\end{table}

\begin{table}
\begin{tabular}{ll}
\textbf{Highest Agency} & \textbf{Lowest Agency} \\
Judge Steven O'Neill       & Kara Swisher \\
Eric Schneiderman          & the United States \\
Russell Simmons            & Hollywood \\
The New York Times         & Meryl Streep \\
Frances McDormand          & \\
CNN                        & \\
Donald Trump               & \\
Hillary Clinton            & \\
\end{tabular}
\caption{Entities with the most the agency and the least agency consist of a mix of men, women, and abstract institutions.}
\label{tab:agency_all}
\end{table}

When we examine power (\Tref{tab:power_all}), the most powerful entities include abstract concepts, like ``the \#MeToo movement'' and ``Twitter''. Women are conspicuously absent from the list of high-powered entities. Instead we find men, including ones directly accused of sexual misconduct (Eric Schniederman). In contrast, women dominate the list of lowest powered entities. While \Tref{tab:power_all} only shows proper nouns, we also observed that common noun references to women were among the least powerful entities identified (e.g. ``a women'', ``these women'').

The agency portrayals are more balanced (Table \ref{tab:agency_all}). While Eric Schneiderman and Donald Trump appear among the entities with highest agency, we also find female supporters of the movement: Frances McDormand and Hillary Clinton.

\begin{figure}[ht]
  \includegraphics[width=\linewidth]{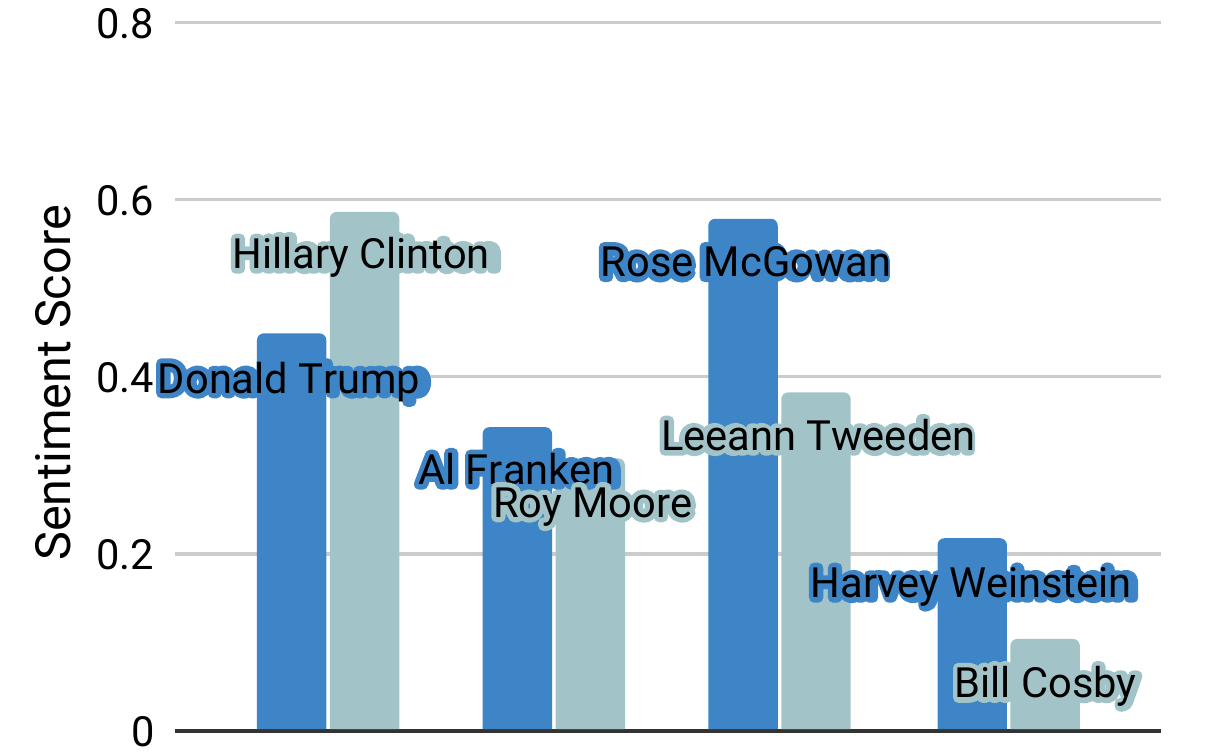}
  \caption{Entities in similar roles are portrayed with different levels of sentiment.}
  \label{fig:sentiment}
    \includegraphics[width=\linewidth]{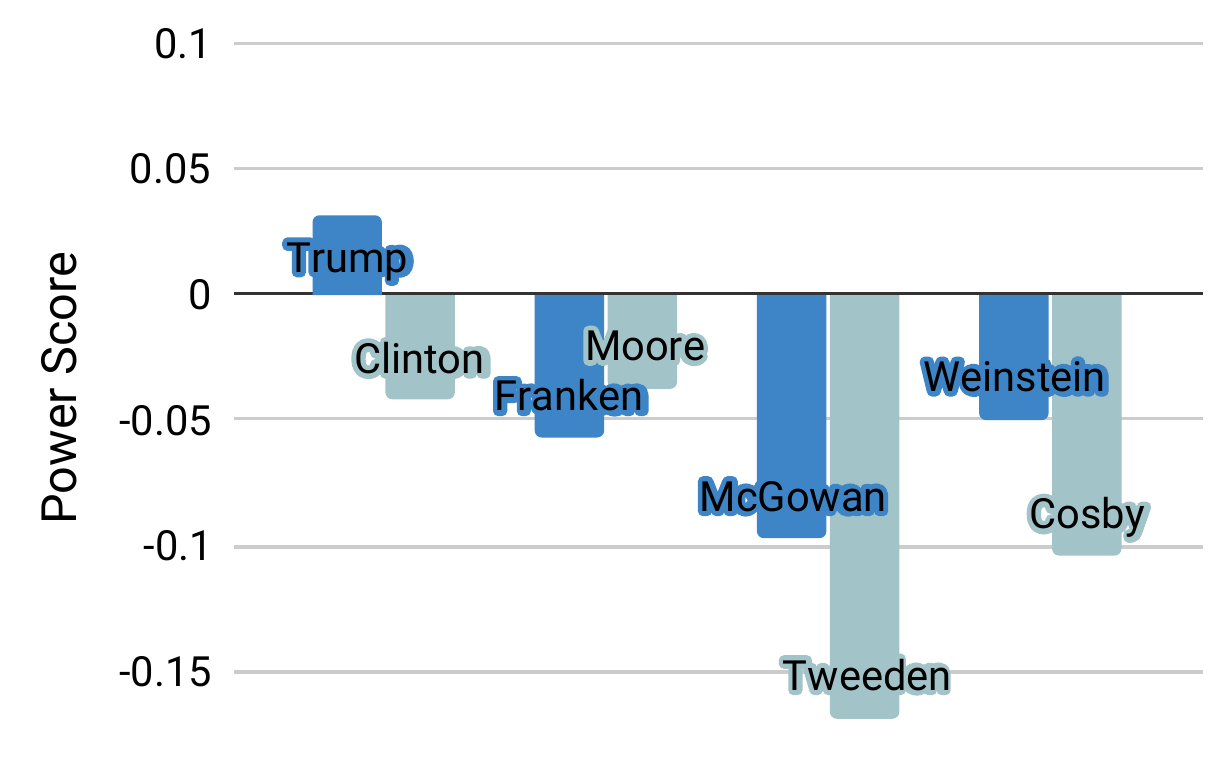}
  \caption{Power scores for comparable entities do not coincide with sentiment scores.}
  \label{fig:power}
\end{figure}

\subsection{Role-Level}

We next conduct a pairwise analysis, where we directly compare two entities who occupied similar roles in different incidents. Through this analysis, we can identify different ways in which narratives of sexual harassment can be framed. Furthermore, by decomposing the pair-wise analysis across different outlets, we can identify bias. How do different outlets cover comparable entities differently?


We directly compare sentiment (Figure \ref{fig:sentiment}) and power (Figure \ref{fig:power}) for several entity pairs. There is a striking difference between the portrayals of Rose McGowan and Leeann Tweeden, two women who both accused high-profile men of sexual assault (Harvey Weinstein and Al Franken respectively). Both women are portrayed with lower power than the men they accused, but Rose McGowan is portrayed with both more positive sentiment and higher power than Leeann Tweeden.
 
Sample articles about Leeann Tweeden focus on accounts of what happened to her: ``The first women to speak out was Leeann Tweeden who said that Franken forcibly kissed her.''\footnote{\url{https://dailym.ai/2WBiA38}} In contrast, news articles about Rose McGowan focus on statements she made after the fact: ``As Rose McGowan, one of the heroes to emerge from the Harvey Weinstein fallout, has mentioned...''.\footnote{\url{https://bit.ly/2TNjtE4}} We can generalize that in this corpus, Rose McGowan matches more of ``survivor'' frame, connoting someone who is proactively fighting, while Leeann Tweeden matches more of a ``victim'' frame, which connotes helplessness and pity.

Figures~\ref{fig:sentiment} and~\ref{fig:power} reveal further differences in portrayals. Democrats Hillary Clinton and Al Franken are portrayed more positively than corresponding Republicans Donald Trump and Roy Moore. However, Donald Trump appears as more powerful than every other entity, which coincides with his role as the current U.S. President. In general, politicians accused of sexual harassment (Roy Moore and Al Franken) are portrayed more positively than entertainment industry figures (Harvey Weinsten and Bill Cosby). While few defended entertainment industry figures accused of harassment, politicians received more mixed support and criticism from their own parties. For instance, Donald Trump publicly endorsed candidate Roy Moore, despite the allegations against him.

\begin{figure}
  \includegraphics[width=\linewidth]{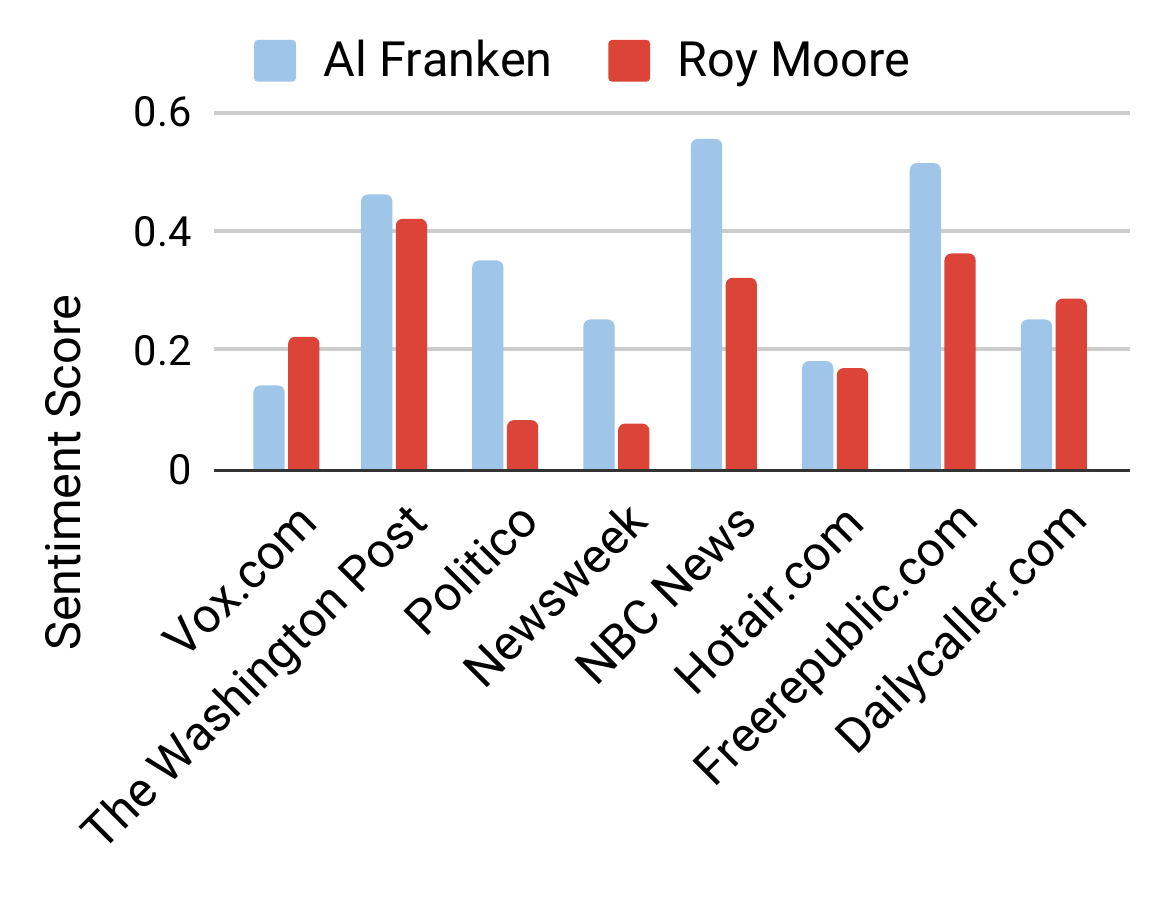}
  \caption{Sentiment scores across left-leaning and right leaning outlets for Democrat Al Franken and Republican Roy Moore do not fall along party lines.\\ \textbf{Left-leaning (Democratic) outlets}: Vox.com,  The Washington Post, Newsweek, NBC News. \\ 
  \textbf{Right-leaning (Republican) outlets}: Hotair.com, Freerepublic.com, Dailycaller.com. \\
  \textbf{Centrist}: Politico\footnote{\url{https://www.allsides.com/}} \\
 }
  \label{fig:franken_moore}
\end{figure}

However, comparing coverage of individuals across the corpus as a whole reveals limited information because it is difficult to separate fact from bias. Does Al Franken receive a more positive portrayal than Roy Moore because his actions throughout the movement were more sympathetic? Or does he receive a more positive portrayal because of a liberal media bias? We can better assess the impact of media bias by comparing how the same entity is portrayed across different outlets. Figure~\ref{fig:franken_moore} shows sentiment scores for Al Franken and Roy Moore across all outlets that mention both entities at least 10 times.

The coverage of Republican Roy Moore falls broadly along party lines, with the lowest-scoring portrayals occurring in left-leaning outlets Politico and Newsweek. Similarly, the left-leaning outlets (with the exception of \url{Vox.com}) portray Democrat Al Franken more positively than Roy Moore. However, the right-leaning outlets portray both men similarly, notably \url{Freerepublic.com} portrays Al Franken more positively than Roy Moore. In reading articles from these outlets, many are sympathetic towards Al Franken, presenting him as a scapegoat, forced out of office by other Democrats without a fair ethics hearing.\footnote{\url{https://bit.ly/2CCjNAW}, \url{https://bit.ly/2COQdao}}  Our analysis reveals surprising differences in how conservative and liberal outlets react differently to events of the \#MeToo movement. Entity portrayals do not necessarily fall along party lines (i.e. liberal outlets portraying liberal politicians positively), but these outlets do focus on different aspects of events.

\begin{figure}
  \includegraphics[width=\linewidth]{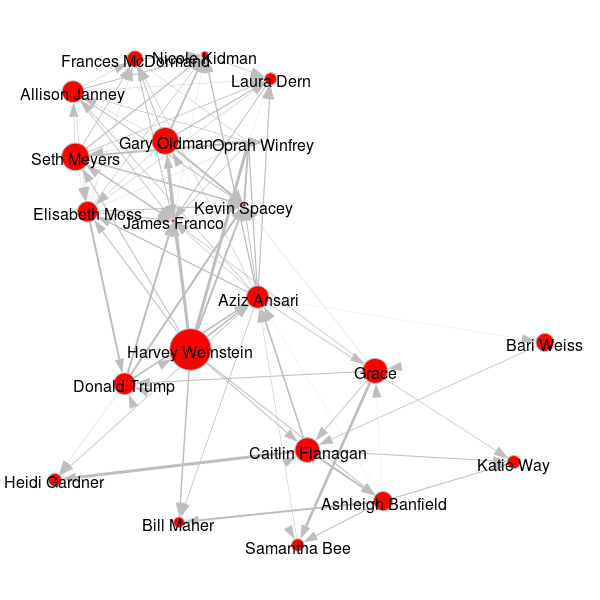}
  \caption{Graphical visualization of power dynamics between entities involved in the accusations against Aziz Ansari}
  \label{fig:aziz_graph}
 \end{figure}

\subsection{Incident-Level}

For the incident-level analysis, we return to the example introduced in the beginning of this paper: the accusations against comedian Aziz Ansari.
First, we examine the power landscape surrounding Aziz Ansari through a graphical visualization. We develop this graph by drawing from psychology theories of power, which suggest that a person's power often derives from the people around them \cite{raven}. We devise a metric to capture the concept of relative power: for any pair of entities, we average the difference between their power scores across all articles in which they are both mentioned. We visualize these differentials in a graph: an edge between two entities denotes that there is at least 1 article that mentions both entities. An edge from entity A to entity B denotes that on average, A is presented as more powerful than B. The magnitude of that difference is reflected in the edge weight. Finally, we sum all of the edge weights to obtain a score for each node. Thus, a large node for entity A indicates that entity A is often portrayed as more powerful than other entities mentioned in the same articles as A.

\begin{figure}
  \includegraphics[width=.9\linewidth]{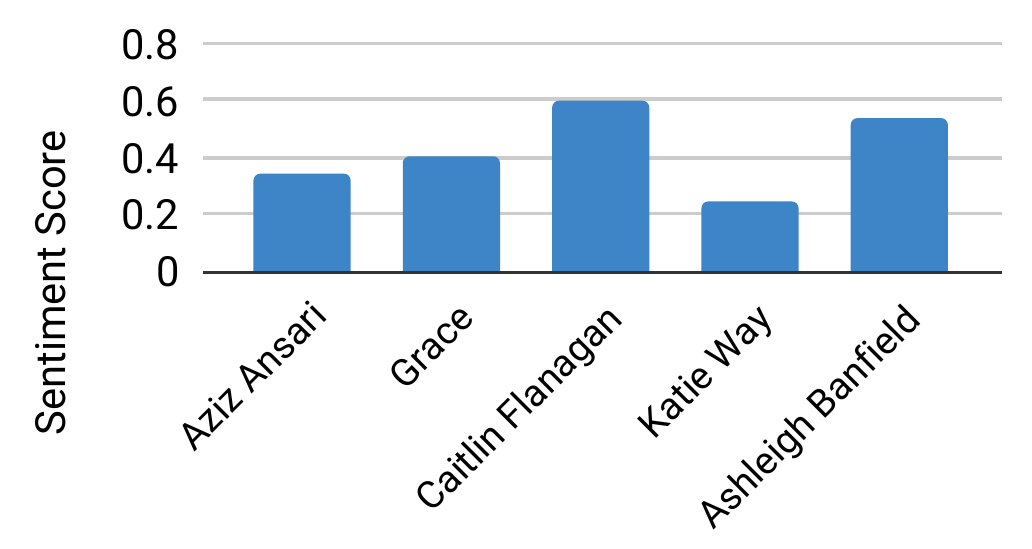}
  \caption{Aziz Ansari and his accuser, Grace, are portrayed with comparable sentiment.}
  \label{fig:aziz_sent}
  
  \includegraphics[width=.9\linewidth]{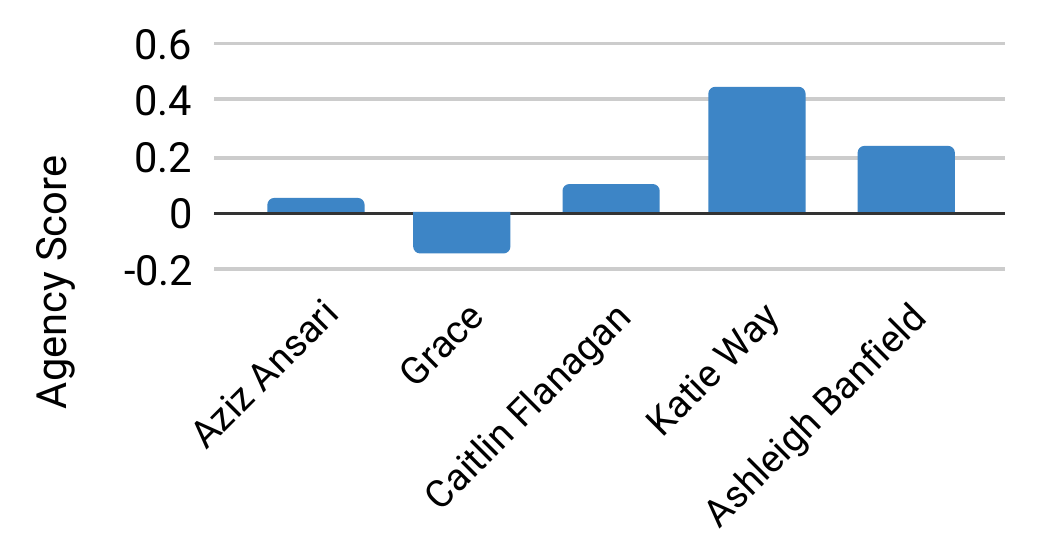}
  \caption{Grace is portrayed with low agency, while journalist Katie Way has very high agency.}
  \label{fig:aziz_agency}

  \includegraphics[width=.9\linewidth]{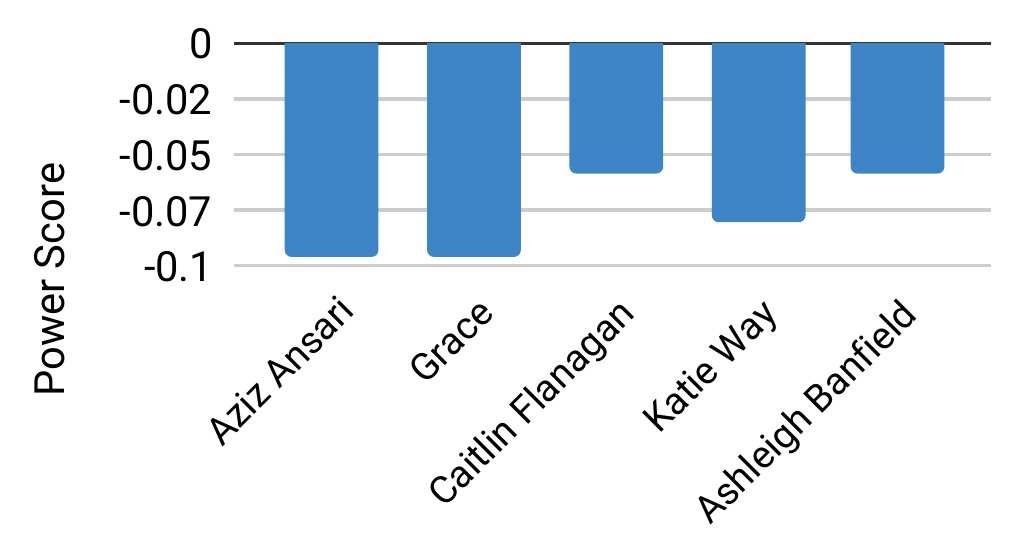}
  \caption{Aziz Ansari and Grace are portrayed with lower power than journalist Ashleigh Banfield.}
  \label{fig:aziz_power}
\end{figure}

\begin{figure}
  \includegraphics[width=.9\linewidth]{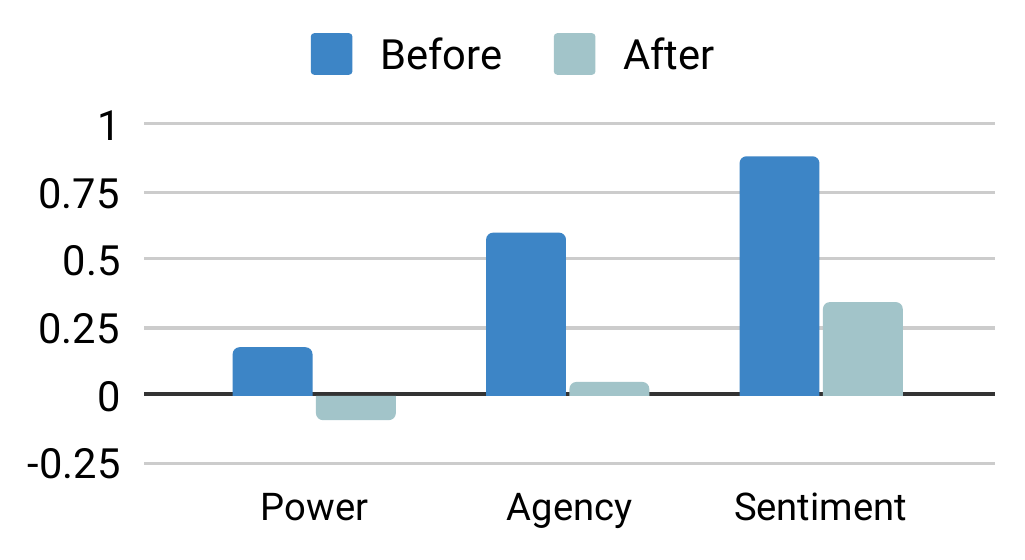}
  \caption{Aziz Ansari is portrayed with lower power, agency, and sentiment in articles published after the allegations against him.}
  \label{fig:aziz_comp}
\end{figure}

 We show this graph for articles related to Aziz Ansari by limiting our corpus to articles that mention ``Aziz'' (555 articles). We take only the 100 most frequently mentioned entities, eliminate all non-people (i.e. Hollywood), and manually group redundant entities (i.e. Donald Trump vs. President Donald Trump). From this graph, we can see two separate clusters, related to two distinct narratives about Aziz Ansari. The cluster of entities in the top left corner, including Seth Meyers, Oprah Winfrey, and Nicole Kidman, is tied to media coverage of the 2018 Golden Globe Awards. All of theses entities won awards or gave speeches, including Aziz Ansari, who won Best Actor in a Television Series -- Musical or Comedy. Much of the Golden Globes centered on the \#MeToo movement, as presenters expressed their support of the movement in speeches or wore pins indicating their support. Aziz Ansari himself wore a ``Time's Up'' pin, to express his opposition to sexual harassment. Thus, because Aziz Ansari won a prominent award and much of the Golden Globes centered on the \#MeToo movement, selecting all articles about Aziz Ansari from our \#MeToo corpus results in many articles about the Golden Globes.
 
In the bottom right corner, we see a second narrative, relating to the accusations against Aziz Ansari. Aziz Ansari and Grace are the primary entities in the narrative. Other frequently mentioned entities include journalists: Katie Way, who authored the original article about the allegations, and Caitlin Flanagan and Ashleigh Banfield, who publicly criticized Katie Way's article. We can see the importance of these journalists in their relative power. For instance, the edge from Caitlin Flanagan to Aziz Ansari indicates that she is often portrayed as more powerful than he is.

We then analyze the sentiment, power, and agency scores for these prominent entities, limiting our data set to articles that mention Aziz Ansari and were published after the \url{Babe.net} article that first disclosed the allegations against him (476 articles). Figure \ref{fig:aziz_sent} portrays sentiment scores. Unlike figures like Harvey Weinstein and Bill Cosby, who have much lower sentiment scores than their accusers, Aziz Ansari has a similar sentiment score to Grace, reflective of the mixed reaction to the article published about them. Katie Way, the journalist who wrote the original article, is portrayed with particularly low sentiment, which coincides with the severe criticism she received for publishing the story. In contrast, Caitlin Flanagan and Ashleigh Banfield, who were both front runners in criticizing Katie Way, were portrayed more positively.

Figure \ref{fig:aziz_agency} shows the agency scores for the same entities. Grace does have lower agency than Aziz Ansari, which supports the critique that Grace lacks any agency in the narrative.  Aziz Ansari and Grace both have lower agency scores in comparison to all 3 journalists. Sentences where Ashleigh Banfield is scored as having high power and agency include: ``Banfield slammed the accuser for her `bad date'.''\footnote{\url{https://bit.ly/2WFVlVz}} The high agency of these journalists demonstrates the prominent role that journalists have in social movements. In contrast Grace's low agency stems from sentences like ``she felt pressured to engage in unwanted sexual acts.'' \footnote{\url{https://bit.ly/2HTvNRx}} The power scores (Figure \ref{fig:aziz_power}) reflect a similar pattern, Grace and Aziz Ansari are portrayed as less powerful than the 3 journalists.

Finally, we compare sentiment, power, and agency scores for Aziz Ansari in articles published before the \url{Babe.net} article (79 articles) and articles published after (476 articles) in Figure \ref{fig:aziz_comp}. As articles about Aziz Ansari before these allegations focus primarily on his Golden Globe victory, these articles portray him with high power, agency, and sentiment. Following the \url{Babe.net} article, we see a decrease in Aziz Ansari's power, agency, and sentiment scores.

\subsection{Limitations and Ethical Implications}
Our analysis is limited by several factors which we identify as areas for future work. First, our metrics for scoring power, agency, and sentiment are based on ternary verb scores.  While our work suggests that verbs are informative, we suspect that there are many other relevant features, including other parts of speech like adjectives and apposition nouns as well as high-level features, like whether or not the entity is directly quoted and how early in the article the entity is first mentioned. Additionally, we assign a ternary (positive, negative, neutral) score to each verb, whereas affect scores are likely best measured with continuous values. The incorporation of additional features and continuous-valued verb scores could provide more accurate entity scores. Our work has also highlighted a need for better evaluation metrics for measuring subjective traits like power. As discussed in \Sref{sec:entity_evaluation}, these tasks are difficult for annotators to perform. Best-Worst Scaling, which has been used for creating affective lexicons \cite{vad-acl2018}, may offer a viable framework. Additionally, our data set consists of an imperfect sample of articles covering the \#MeToo movement and a different sampling of articles may yield different results.

Furthermore, although our goal in analyzing media coverage is to promote positive journalism and advocate for media that empowers underrepresented people, we acknowledge that our work has the potential to be misused. Tools for analyzing media portrayals can be used to intentionally present biased viewpoints and manipulate public opinion. Furthermore, our analysis of actual people could have unforeseen consequences on them and their public images.

\section{Conclusions}

We present contextual affective analysis: a framework for analyzing nuanced entity portrayals. While we focus specifically on power, agency, and sentiment in media coverage of the \#MeToo movement, our approach is a general framework that can be adapted to other corpora and research questions. Our analysis of media coverage of the \#MeToo movement addresses questions like ``Whom does the media portray as sympathetic?'' and ``Whom does the media portray as powerful?'' We demonstrate that although this movement has empowered women by encouraging them to share their stories, this empowerment does not necessarily translate into online media coverage of events. While women are among the most sympathetic entities, traditionally powerful men remain among the most powerful in media reports. We further show the prominence of journalists and 3rd party entities commenting on events without being directly involved, not only because their statements can influence perception of the movement, but because by making statements, they become entities in the narrative. Through this analysis, we highlight the importance of media framing: journalists can choose which narratives to highlight in order to promote certain portrayals of people. They can encourage or undermine movements like the \#MeToo movement through their choice of entity portrayal.

\section*{Acknowledgements}

We gratefully acknowledge Alan Black, Hannah Rashkin, Maarten Sap, Emily Ahn, and our anonymous reviewers for their helpful advice. We further thank our annotators. This material is based upon work supported by the NSF Graduate Research Fellowship Program under Grant No.~DGE1745016 and by Grant No.~IIS1812327 from the NSF. Any opinions, findings, and conclusions or recommendations expressed in this material are those of the authors and do not necessarily reflect the views of the NSF.

\section*{Appendix}

An alphabetical list of people whom we refer to in connection with the \#MeToo movement:

\noindent \textbf{Woody Allen} - U.S. movie director; repeatedly accused of sexual assault by daughter Dylan Farrow since 2013 \\
\textbf{Aziz Ansari} - U.S. comedian; accused of sexual misconduct by an anonymous woman in January 2018 \\
\textbf{Ashleigh Banfield} - Canadian-American journalist; criticized the article levelling accusations of sexual assault against Aziz Ansari in an open letter \\
\textbf{Tarana Burke} - U.S. civil rights activist; started the \#MeToo movement to raise awareness about sexual assault \\
\textbf{Hillary Clinton} - Democratic Party's nominee for President of the United States in the 2016 election, former U.S. Senator and Secretary of State \\
\textbf{Andrea Constand} - Accused Bill Cosby of sexually assaulting her in 2004; though many women accused Cosby of harassment, only Constand brought her case to court \\
\textbf{Bill Cosby} - Former comedian; publicly accused of sexual assault or harassment by 60 women; found guilty of assaulting Andrea  Constand in April 2018 \\
\textbf{Caitlin Flanagan} - U.S. writer; wrote a widely-circulated article that was critical of the accusations of sexual misconduct against Aziz Ansari \\
\textbf{Al Franken} - Former U.S. senator; resigned in December 2017 following several allegations of sexual misconduct \\
\textbf{Grace} - pseudonym employed by the woman whose accusations of sexual misconduct against Aziz Ansari were published in a \url{Babe.net} article \\
\textbf{Jimmy Kimmel} - U.S. television host; delivered a widely reported speech in support of the \#MeToo movement as the host of the Oscars in March 2018 \\
\textbf{Meghan Markle} - Former U.S. actress; married Prince Harry in May 2018 (with much media coverage of engagement leading up to the wedding); public supporter of \#MeToo movement \\
\textbf{Frances McDormand} - U.S. actress; prominent because of her support of \#MeToo movement during the 2018 Oscars  \\
\textbf{Rose McGowan} - U.S. actress; one of the first women to accuse Harvey Weinstein of sexual misconduct in Oct 2017 \\
\textbf{Alyssa Milano} - U.S. actress; posted a tweet with the hashtag \#MeToo and encouraged others who had been sexually harassed to do the same \\
\textbf{Roy Moore} - U.S. politician; accused by multiple women of sexual misconduct; ran (and lost) for the U.S. Senate in 2017 and was publicly endorsed by Donald Trump \\
\textbf{Judge Steven O'Neill} - Montgomery Country judge; sentenced Bill Cosby to prison and labeled him a sex offender \\
\textbf{Eric Schneiderman} - Former New York State Attorney General; resigned in May 2018 following accusations of sexual misconduct by four women \\
\textbf{Ryan Seacrest} - U.S. television personality and producer; accused of sexual misconduct by his stylist in November 2017 \\
\textbf{Russel Simmons} - U.S. record producer; accused by 18 women of sexual assault \\
\textbf{Meryl Streep} - U.S. actress who has been accused of long knowing about, but not condemning, Weinstein's alleged sexual misconduct \\
\textbf{Kara Swisher} - U.S. journalist \\
\textbf{Uma Thurman} - U.S. actress; detailed charges of sexual assault against Harvey Weinstein in February 2018 \\
\textbf{Leeann Tweeden} - U.S. radio broadcaster; accused Al Franken of sexual misconduct in November 2017 \\
\textbf{Katie Way} - author of an article published at \url{Babe.net} which accused Aziz Ansari of sexual misconduct against anonymous `Grace', and which set off a public dialogue about consent and the definition of sexual misconduct \\
\textbf{Harvey Weinstein} - Former film producer, accused by over 80 women of sexual misconduct; accusations against him sparked the \#MeToo movement \\
\textbf{Oprah Winfrey} - U.S. media executive and television personality; gave a widely reported speech in support of the \#MeToo movement at the Golden Globes in January 2018 \\

\bibliography{references.bib}
\bibliographystyle{aaai}

\end{document}